\documentclass[11pt,ams]{article}%
\usepackage{geometry}
\usepackage{dsfont}
\usepackage{amsmath}
\usepackage{amsfonts}
\usepackage{amssymb}
\usepackage{graphicx}%

\newcommand{\p}{\partial}

\newcommand{\s}{\sigma}

\renewcommand{\d}{\delta}

\newcommand{\e}{\varepsilon}

\begin{document}

\date{}
\title{\textbf{Study of  the zero modes of the Faddeev-Popov operator in the maximal Abelian gauge}}
\author{\textbf{M.~A.~L.~Capri}\thanks{caprimarcio@gmail.com}\,\,,
\textbf{M.~S.~Guimaraes}\thanks{msguimaraes@uerj.br}\,\,,
\textbf{V.~E.~R.~Lemes}\thanks{vitor@dft.if.uerj.br}\,\,,
\textbf{S.~P.~Sorella}\thanks{sorella@uerj.br}\,\,\thanks{Work supported by FAPERJ, Funda\c{c}\~{a}o de Amparo \`{a} Pesquisa do Estado do Rio de Janeiro, under the program \textit{Cientista do Nosso Estado}, E-26/101.578/2010}\,\,,\\
\textbf{D.~G.~Tedesco}\thanks{dgtedesco@uerj.br}\\
[2mm]
\textit{{\small{UERJ $-$ Universidade do Estado do Rio de Janeiro}}}\\
\textit{{\small{Instituto de F\'{\i}sica $-$ Departamento de F\'{\i}sica Te\'{o}rica}}}\\
\textit{{\small{Rua S\~{a}o Francisco Xavier 524, 20550-013 Maracan\~{a}, Rio de Janeiro, RJ, Brasil.}}}
}
\maketitle
\begin{abstract}
A study of the zero modes of the Faddeev-Popov operator in the maximal Abelian gauge is presented in the case of the gauge group  $SU(2)$ and for different Euclidean space-time dimensions. Explicit examples of classes of normalizable zero modes and corresponding gauge field configurations are constructed by taking into account two boundary conditions, namely:  $i)$ the finite Euclidean Yang-Mills action, $ii)$ the finite Hilbert norm.
\end{abstract}
\section{Introduction}
The gauge fixing of a non-abelian gauge theory is one of the main ingredient in order to achieve a quantization procedure within a local, covariant and renormalizable  quantum field theory framework. The usual method employed to fix the gauge  is the Faddeev-Popov procedure. This procedure is, however, valid only at the perturbative domain, due to the existence of the so-called Gribov copies  \cite{Gribov:1977wm}, {\it i.e.} of equivalent field configurations fulfilling the same gauge condition. \\\\As shown by Singer \cite{Singer:1978dk}, the existence of the Gribov copies is  a generic feature of the gauge fixing. Gribov copies have in fact been detected in all commonly employed local, covariant and renormalizable gauge fixings. As pointed out in  \cite{Singer:1978dk}, the Gribov problem is a non-perturbative obstruction, characterizing the impossibility of selecting a unique gauge configuration for each gauge orbit by means of a local and covariant gauge condition. \\\\Examples of Gribov copies can be obtained by looking at the zero modes of the Faddeev-Popov operator corresponding to a given gauge-fixing condition. The presence of zero modes makes the Faddeev-Popov procedure ill-defined. As argued by Gribov in his seminal work \cite{Gribov:1977wm}, a suitable restriction of the domain of integration in the  functional integralto the so called Gribov region $\Omega$ \cite{Gribov:1977wm,Vandersickel:2012tz,Sobreiro:2005ec}  is needed, resulting in a  deep modification of the behavior of the correlation functions of the theory in the nonperturbative infrared region, see \cite{Vandersickel:2012tz} for a recent account on the Gribov issue. \\\\The present work deals with the study of the zero modes of the Faddeev-Popov operator in the maximal Abelian gauge \cite{Bruckmann:2000xd} in various Euclidean space-time dimensions and  for the gauge group $SU(2)$. We shall provide explicit examples of classes of normalizable zero modes and associated gauge configurations by taking into account two boundary conditions for the gauge fields, namely: $i)$ the finite Euclidean Yang-Mills action, $S_{\mathrm{YM}}=\frac{1}{4g^{2}}\int d^{d}x\,F^{a}_{\mu\nu}F^{a}_{\mu\nu} < \infty$, $ii)$ the finite Hilbert norm, $\|A\|^{2}=\int d^{d}x\, A^{a}_{\mu}A^{a}_{\mu}< \infty$. The choice of these boundary conditions has a rather clear physical meaning. The requirement of finite action is equivalent to that of finite energy, as the Yang-Mills action enters  the Boltzmann weight of the Euclidean functional integral. On the other hand, the Hilbert norm  $\|A\|^{2}$ is known to be deeply related to the Gribov issue. In fact, several properties of the Gribov region in both Landau and maximal Abelian gauge can be obtained by looking at all relative minima of  $\|A\|^{2}$ along the gauge orbits, see \cite{Vandersickel:2012tz,Sobreiro:2005ec,Capri:2005tj,Capri:2006cz,Capri:2008ak,Capri:2008vk,Capri:2010an}.\\\\This work generalizes to the case of the maximal Abelian gauge  the results obtained in \cite{Guimaraes:2011sf, Capri:2012ev} for the Landau gauge,   where  classes of normalizable zero modes have been obtained. As done in \cite{Guimaraes:2011sf, Capri:2012ev}, we shall make use of  Henyey's construction \cite{Henyey:1978qd}, which allows for a self-consistent characterization of the zero modes and corresponding gauge field configurations\footnote{For a review of Henyey's method, see \cite{Sobreiro:2005ec,Guimaraes:2011sf}. }.  More precisely, the differential equation for the zero mode is employed as an algebraic equation allowing us to express the gauge field  in terms of the function which parametrizes the zero mode itself. This self-consistent construction can be summarized by the following steps: $a)$ an ansatz  for the gauge field configuration fulfilling the gauge  condition is written down,  $b)$ a class of normalizable solutions corresponding to zero modes of the Faddeev-Popov is proposed. At this stage, the differential equation for the zero modes is employed as an algebraic equation to determine the corresponding gauge configurations in terms of the functions parametrizing the zero modes,  $c$) one finally checks if the resulting gauge fields  fulfill the boundary condition which has been imposed at infinity.  \\\\The  work is organized as follows. In Section 2 we review the Faddeev-Popov quantization procedure in  $SU(2)$ Yang-Mills theory in the maximal Abelian gauge. In Section 3 we construct zero modes solutions of the Faddeev-Popov operator in $d=2$ Euclidean dimensions. In this case, we shall be able to obtain a class of normalizable zero modes with corresponding gauge field configurations displaying finite action but not finite Hilbert norm. In Section 4 we face the case of $d=3$ Euclidean dimensions. Here, we shall obtain a normalizable zero modes solutions with corresponding gauge fields configurations displaying finite action as well as  finite norm. In Section 5 we discuss the $d=4$ case. Also here,  we shall be able to construct normalizable  zero modes  with corresponding gauge fields configurations fulfilling both boundary conditions. Finally, in Section 6 we present our conclusions.

\section{A brief review of the Faddeev-Popov quantization in the maximal Abelian gauge}

\subsection{The gauge fixing conditions}
Let us first remind some properties of the maximal Abelian gauge for the gauge group $SU(2)$. The gauge field $\mathcal{A}_{\mu}(x)$ can be decomposed into its off-diagonal and diagonal components, as follows
\begin{equation}
\mathcal{A}_{\mu}=A^{a}_{\mu}T^{a}+A_{\mu}T^{3}\,,
\end{equation}
where $T^{a}$, with $a=1,2$, are the off-diagonal generators of $SU(2)$, while $T^{3}$ stands for the diagonal one. The three generators are given in terms of Pauli matrices, $T^{i}=\sigma^{i}/2$ $(i=1,2,3)$, and obey the following commutation relations: 
\begin{equation}
\Bigl[T^{a},T^{b}\Bigr]=i\e^{ab}T^{3}\,,\qquad
\Bigl[T^{3},T^{a}\Bigr]=i\e^{ab}T^{b}\,,
\end{equation}
where $\e^{ab}=-\e^{ba}\equiv\e^{ab3}$ is totally anti-symmetric: 
\begin{equation}
\e^{ac}\e^{bd}=\d^{ab}\d^{cd}-\d^{ad}\d^{cb}\,.
\label{epsilon_product}
\end{equation}
Thus, the Yang-Mills action in Euclidean $d$-dimensional Euclidean space-time  can be written as
\begin{equation}
S_{\mathrm{YM}}=\frac{1}{4g^{2}}\int d^{d}x\,\left(F^{a}_{\mu\nu}F^{a}_{\mu\nu}+F_{\mu\nu}F_{\mu\nu}\right)\,.\label{YM_action}
\end{equation}
Here, the off-diagonal and diagonal  components of the field strength are given by:
\begin{eqnarray}
F^{a}_{\mu\nu}&=&D^{ab}_{\mu}A^{b}_{\nu} - D^{ab}_{\nu}A^{b}_{\mu}\,,\nonumber\\
F_{\mu\nu}&=&\p_{\mu}A_{\nu} - \p_{\nu}A_{\mu} +\e^{ab}A^{a}_{\mu}A^{b}_{\nu}\,,
\end{eqnarray}
with $D^{ab}_{\mu}$ being the covariant derivative with respect to the diagonal components, $A^3_\mu \equiv A_\mu$,
\begin{equation}
D^{ab}_{\mu}=\d^{ab}\p_{\mu} - \e^{ab}A_{\mu}\,.
\label{cov}
\end{equation}
The action \eqref{YM_action} is left invariant by the gauge transformations:
\begin{eqnarray}
\d A^{a}_{\mu}&=& -D^{ab}_{\mu}\omega^{b}-\e^{ab}A^{b}_{\mu}\omega\,,\nonumber\\
\d A_{\mu}&=&-\p_{\mu}\omega-\e^{ab}A^{a}_{\mu}\omega^{b}\,,\label{gauge_transf}
\end{eqnarray}
where $(\omega^{a},\omega)$ stand for the off-diagonal and diagonal components of the gauge parameter.\\\\ The gauge fixing conditions defining the maximal Abelian gauge are:
\begin{equation}
D^{ab}_{\mu}A^{b}_{\mu}=0\,,\qquad \p_{\mu}A_{\mu}=0\,.\label{gauge_fixing}
\end{equation}
Notice that the condition imposed on the off-diagonal components is non-linear and it can be obtained by demanding that the auxiliary functional,
\begin{equation}
\|A\|^{2}=\int d^{d}x\, A^{a}_{\mu}A^{a}_{\mu}\,,
\end{equation}
which corresponds to the norm of the off-diagonal components,  is stationary  with respect to the gauge transformations \eqref{gauge_transf}.
Moreover, this condition allows for a residual local $U(1)$ invariance
corresponding to the diagonal subgroup of $SU(2)$. This additional invariance has to be fixed
by means of a suitable gauge condition on the diagonal component $A_{\mu}$. It is common to choose a Landau type condition, such as written in eq.\eqref{gauge_fixing}, also adopted in lattice simulations.\\\\The gauge fixing conditions for the maximal Abelian gauge, eq.\eqref{gauge_fixing}, can be implemented by means of the Faddeev-Popov quantization method, which is expressed by the following partition function:
\begin{eqnarray}
\mathcal{Z}=\int \mathcal{D}A^{a}_{\mu}\mathcal{D}A_{\mu}\,\mathcal{D}\bar{c}^{a}\mathcal{D}{c}^{a}\,\mathcal{D}\bar{c}\mathcal{D}{c}\,
\mathcal{D}{b}^{a}\mathcal{D}{b}\,\,e^{-S_{\mathrm{FP}}[A,b,\bar{c},c]}\,,
\end{eqnarray}
where the action $S_{\mathrm{FP}}$ is given by
\begin{equation}
S_{\mathrm{FP}}=S_{\mathrm{YM}}+S_{\mathrm{MAG}}\,,
\end{equation} 
with $S_{\mathrm{YM}}$ being given by \eqref{YM_action} and
\begin{equation}
S_{\mathrm{MAG}}=\int d^{d}x\,\Bigl[ib^{a}\,D^{ab}_{\mu}A^{b}_{\mu} - \bar{c}^{a}\mathcal{M}^{ab}c^{b} -\e^{ab}(D^{ac}_{\mu}A^{c}_{\mu})\bar{c}^{b}c
+ib\,\p_{\mu}A_{\mu}+\bar{c}^{a}\,\p_{\mu}(\p_{\mu}c+\e^{ab}A^{a}_{\mu}c^{b})\Bigr]\,.
\end{equation}
The fields $(b^{a},b)$ denote the Lagrange multipliers enforcing the gauge conditions, while the fields $(\bar{c}^{a},c^{a})$ and $(\bar{c},c)$ are the off-diagonal and diagonal components of the Faddeev-Popov ghosts. The operator $\mathcal{M}^{ab}$ is the Faddev-Popov operator and  is given by:
\begin{equation}
\mathcal{M}^{ab}=-D^{ac}_{\mu}D^{cb}_{\mu}-\e^{ac}\e^{bd}A^{c}_{\mu}A^{d}_{\mu}\,.
\label{FPop}
\end{equation}
As expected, $\mathcal{M}^{ab}$ is non-linear\footnote{In the case of the Landau gauge, the corresponding Faddeev-Popov operator is linear in the gauge fields, {\it i.e.}  
\begin{equation}
\mathcal{M}^{ab}_{\rm Landau} = -\delta^{ab} \partial^2 + gf^{abc} A^c_\mu\partial_\mu \;, \qquad {\rm with} \qquad a,b,c = 1,...., N^2-1 \;, 
\end{equation}
where $f^{abc}$ are the structure constants of $SU(N)$. }
in the gauge fields, due to the non-linearity of the gauge condition \eqref{gauge_fixing}.  \\\\The action $S_{\mathrm{FP}}$ is left invariant by the nilpotent BRST transformations
\begin{eqnarray}
sA^{a}_{\mu}&=&-(D^{ab}_{\mu}c^{b}+\e^{ab}A^{b}_{\mu}c)\,,\nonumber\\
sA_{\mu}&=&-(\p_{\mu}c+\e^{ab}A^{a}_{\mu}c^{b})\,,\nonumber\\
sc^{a}&=&\e^{ab}c^{b}c\,,\nonumber\\
sc&=&\frac{1}{2}\e^{ab}c^{a}c^{b}\,,\nonumber\\
s\bar{c}^{a}&=&ib^{a}\,,\nonumber\\
sb^{a}&=&0\,,\nonumber\\
s\bar{c}&=&ib\,,\nonumber\\
sb&=&0\,,\label{brst}
\end{eqnarray} 
which enable us to prove that the maximal Abelian gauge is a renormalizable gauge \cite{Min:1985bx,Fazio:2001rm}. \\\\The Lagrange multipliers $(b^{a},b)$ can be integrated out and the diagonal ghosts $(\bar{c},c)$ can be completely decoupled from the theory by means of the following  change of variables 
\begin{equation}
c\to\xi=c+\e^{ab}\frac{\p_{\mu}}{\p^{2}}A^{a}_{\mu}c^{b}\,,\qquad\bar{c}\to\bar{\xi}=\bar{c}\,,
\end{equation}
with all other fields unchanged. Being linear in the fields $(\bar{c},c)$  this change of variables leads to a Jacobian which
is field independent. Thus, one verifies that
\begin{equation}
\bar{c}\,\p_{\mu}(\p_{\mu}c+\e^{ab}A^{a}_{\mu}c^{b})\to\bar{\xi}\,\p^{2}\xi \;. 
\end{equation}
For the partition function one gets
\begin{equation}
\mathcal{Z}=\mathcal{N}\int\mathcal{D}A^{a}_{\mu}\mathcal{D}A_{\mu}\,\mathcal{D}\bar{c}^{a}\mathcal{D}c^{a}\,
\d(D^{ab}_{\mu}A^{b}_{\mu})\d(\p_{\mu}A_{\mu})
\,e^{-S_{\mathrm{YM}}+\int d^{d}x\,\bar{c}^{a}\mathcal{M}^{ab}c^{b}}\,,
\end{equation}
where $\mathcal{N}$ in an irrelevant normalization factor. Finally, integrating out the off-diagonal ghosts, we obtain
\begin{equation}
\mathcal{Z}=\mathcal{N}\int\mathcal{D}A^{a}_{\mu}\mathcal{D}A_{\mu}\,
\d(D^{ab}_{\mu}A^{b}_{\mu})\d(\p_{\mu}A_{\mu})\det(\mathcal{M}^{ab})\,
\,e^{-S_{\mathrm{YM}}}\,.  \label{pfmag}
\end{equation}   
Expression \eqref{pfmag} is the starting point in order to investigate the non-perturbative effects related to the existence of the Gribov copies. Although being out of the aim of the present work, it is worth mentioning that a treatment of the Gribov issue similar to that of the Landau gauge  \cite{Vandersickel:2012tz,Dudal:2008sp} can be implemented in the maximal Abelian gauge, see for instance refs.\cite{Capri:2005tj,Capri:2006cz,Capri:2008ak,Capri:2008vk,Capri:2010an}. Let us turn thus to the characterization of the zero modes of the Faddeev-Popov operator $\mathcal{M}^{ab}$, which clearly affect expression \eqref{pfmag}.

\section{Constructing zero modes in $d=2$ Euclidean space-time}

We would like to solve the equation
\begin{equation}
\mathcal{M}^{ab}\omega^{b}=0\,,
\label{zero_mode_eq}
\end{equation}
where $\mathcal{M}^{ab}$ is the Faddeev-Popov operator, given by eq.\eqref{FPop}, which can be written in a convenient way as
\begin{equation}
\mathcal{M}^{ab}
=-\p^{2}\d^{ab}+2\e^{ab}A_{\mu}\p_{\mu}+A_{\mu}A_{\mu}\d^{ab}
-A^{c}_{\mu}A^{c}_{\mu}\d^{ab}+A^{a}_{\mu}A^{b}_{\mu}\,,
\end{equation}
where use has been made of the eqs.\eqref{epsilon_product} and \eqref{cov}, and of the transversality of the diagonal component of the gauge field, $\partial_\mu A_\mu=0$. Being the maximal Abelian gauge conditions given by eq.\eqref{gauge_fixing}, we will choose a field configuration which automatically fulfills these conditions, namely  
\begin{equation}
A_{\mu}=0\,,\qquad
A^{a}_{\mu}=\d^{a1}\e_{\mu\nu}x_{\nu}f(r)\,,
\label{field_config_2d}
\end{equation}
where $f(r)$ is a spherical symmetric function. In other words, we have for the off-diagonal components the following expressions
\begin{equation}
A^{1}_{1}=yf(r)\,,\qquad A^{1}_{2}=-xf(r)\,,\qquad A^{2}_{\mu}=0\;. \label{f}
\end{equation}
Substituting  eqs.\eqref{f} in the zero-mode equation \eqref{zero_mode_eq}, we obtain:
\begin{eqnarray}
\p^{2}\omega_{1}&=&0\,,\nonumber\\
\p^{2}\omega_2+ r^2f^2(r)\omega_2&=&0\,.\label{sistem}
\end{eqnarray}
The first equation of \eqref{sistem} implies that $\omega_1$ vanishes, $\omega_1=0$, due to the asymptotic behavior at infinity required in order to have a normalizable zero mode.  We are left then with the second equation of \eqref{sistem}.\\\\Reminding that he Laplacian operator in two-dimensional polar coordinates is given by:
\begin{equation}
\partial^{2}\equiv\nabla^{2}=\frac{1}{r}\frac{\partial}{\partial r}\left(r\frac{\partial}{\partial r}\right)
+\frac{1}{r^{2}}\frac{\partial^{2}}{\partial\varphi^{2}}\,,
\end{equation}
it follows that  eq.\eqref{sistem} takes the form
\begin{equation}
\frac{1}{r}\frac{\partial}{\partial r}\left(r\frac{\partial\omega_2}{\partial r}\right)
+\frac{1}{r^{2}}\frac{\partial^{2}\omega_2}{\partial\varphi^{2}}+r^2f^2(r)\omega_2=0\,.
\end{equation}
Applying the method of the separation of variables, we write $\omega_2$ as
\begin{equation}
\omega_2(r,\varphi)=\sigma(r)\xi(\varphi)\,.  \label{zm}
\end{equation}
Thus,
\begin{equation}
\frac{r}{\sigma}\frac{d}{dr}\left(r\frac{d\sigma}{dr}\right)+r^{4}f^{2}(r)=
-\frac{1}{\xi}\frac{d^{2}\xi}{d\varphi^{2}}=k\equiv \mathrm{constant}\,,
\end{equation}
{\it i.e.} we have two ordinary differential equations:
\begin{eqnarray}
\frac{d^{2}\xi}{d\varphi^{2}}+k\xi&=&0\label{xi_eq}\,,\\
\frac{d}{dr}\left(r\frac{d\sigma}{dr}\right)+r^{3}f^{2}(r)\sigma-\frac{k}{r}\sigma&=&0\label{sigma_eq}\,.
\end{eqnarray}
Equation \eqref{xi_eq} is easily solved and its general solution is
\begin{equation}
\xi(\varphi)=c_1\,e^{i\sqrt{k}\,\varphi}+c_2\,e^{-i\sqrt{k}\,\varphi}\,.
\end{equation}
We can still choose the particular solution
\begin{equation}
c_1=c_2=1/2\,,
\end{equation}
which  will be assumed  from now on, namely
\begin{equation}
\xi(\varphi)=\cos(\sqrt{k}\,\varphi)\,.
\end{equation}
Let us turn to eq.\eqref{sigma_eq} which takes a simpler form by setting 
\begin{equation}
\sigma(r)=r\psi(r)\,,
\end{equation}
so that
\begin{equation}
\psi''+\frac{3}{r}\psi'+\left(r^{2}f^{2}(r)+\frac{1-k}{r^{2}}\right)\psi=0\,. \label{psi2}
\end{equation}
According to Henyey's construction \cite{Henyey:1978qd}, we employ eq.\eqref{psi2} as an algebraic equation expressing the function $f(r)$ determining the gauge field configuration, eqs.\eqref{f}, in terms of the quantity $\psi(r)$ which parametrizes the zero mode itself, eq.\eqref{zm}. Therefore, 
\begin{equation}
r^{2}f^{2}(r)=-\frac{1}{r\psi}\left(r\psi''+3\psi'+\frac{1-k}{r}\psi\right)\,.
\end{equation}
We proceed now by writing down a possible trial expression for $\psi$ which gives rise to a normalizable zero mode $\omega_2$ as well as to a gauge configuration fulfilling the desired boundary conditions. Our first concern is to show that the gauge field configuration is real. Then, for a given $\psi(r)$, we have to prove that the quantity $r^{2}f^{2}(r)$ is positive.\\\\As trial expression for $\psi(r)$ we take
\begin{equation}
\psi(r)=\frac{1}{r^{p}+\beta} \,,
\end{equation}
where $p>0$ and $\beta$ is a positive arbitrary constant. 
The norm of the zero-mode $\omega_2$ turns out to be given by 
\begin{equation}
\parallel\omega_2\parallel^{2}=\int_{0}^{\infty}\int_{0}^{2\pi}rdrd\varphi\,(\cos(\sqrt{k}\,\varphi))^{2}\frac{r^{2}}{(r^{p}+\beta)^{2}}\,,
\end{equation}
which is finite if $p>2$. The quantity $r^{2}f^{2}(r)$ is then written as
\begin{equation}
r^{2}f^{2}(r)=\frac{1}{(r^{p}+\beta)^{2}}\left[r^{2p-2}\,(2p-p^{2}-1+k)
+\beta\,r^{p-2}\,(p^{2}+2p-2+2k)+\beta(k-1)\,r^{-2}\right]\,.    \label{gf2}
\end{equation}
Thus, $r^{2}f^{2}(r)$ will be positive if
\begin{equation}
k\geq1+p(p-2)\,,\qquad\mathrm{and}\qquad p>2\,.
\label{kmrelation}
\end{equation}
Notice also that 
\begin{equation}
f(r)\sim 1/r^{2} \qquad  {\rm as} \;\; r\to \infty \;.    \label{ib}
\end{equation}
Having found the explicit expression for the gauge field, eq.\eqref{gf2}, let us look at the boundary conditions. To that end, let us evaluate the  field strength $F^a_{\mu\nu}$ which, for the field configuration \eqref{field_config_2d},  is given by
\begin{equation}
F^{1}_{\mu\nu}=\partial_{\mu}A^{1}_{\nu}-\partial_{\nu}A^{1}_{\mu}\,,\qquad
F^{2}_{\mu\nu}=F^{3}_{\mu\nu}=0\,.
\end{equation}
As a consequence, the Yang-Mills action is found to be  
\begin{equation}
S_{\mathrm{YM}}=\frac{1}{4g^{2}}\int d^{2}x\, F^{1}_{\mu\nu}F^{1}_{\mu\nu}=
\frac{\pi}{g^{2}}\int_{0}^{\infty}dr\,r\left(2f(r)+rf'(r)\right)^{2}\,. \label{ym2}
\end{equation}
This expression is convergent, due to the behavior of $f(r)$ at infinity, eq.\eqref{ib}. \\\\Let us turn now to the Hilbert norm
\begin{equation}
\|\mathcal{A}\|^{2} =\int d^{2}x\, A^{a}_{\mu}A^{a}_{\mu}=2\pi\int_{0}^{\infty}dr\, r^{3}f^{2}(r)\,. \label{hn2}
\end{equation}
Unlike the Yang-Mills action, expression \eqref{hn2} is logarithmic divergent for any valid value of $p$ (and consequently for $k$), {\it i.e.} for any $p>2$. In summary, we have been able then to obtain a class of normalizable zero-modes of the Faddeev-Popov operator, parametrized by the parameters
$k$ and $p>2$, see  eq.\eqref{kmrelation}, whose corresponding gauge field configuration, \eqref{field_config_2d}, \eqref{gf2}, yields a  finite Yang-Mills action. Though, the Hilbert norm $\|\mathcal{A}\|^{2}$ is logarithmic divergent.

\section{The $d=3$ case}

As starting gauge field configuration we take now
\begin{equation}
A_{\mu}=0\,,\qquad A^{a}_{\mu}=\varepsilon_{a\mu\nu}\,x_{\nu}\,g(r)\,,
\label{field_config_3d}
\end{equation}
with $g(r)$ being a spherical symmetric function and $\mu=1,2,3$. The definition above implies that
\begin{equation}
A^{1}_{1}=0\,,\qquad A^{1}_{2}=-A^{2}_{1}=z\,g(r)\,,\qquad A^{1}_{3}=-y\,g(r)\,,\qquad A^{2}_{2}=0\,,\qquad A^{2}_{3}=x\,g(r)\,.
\end{equation}
It is straightforward to show that this field configuration fulfills the gauge conditions of the maximal Abelian gauge,  eq.\eqref{gauge_fixing}. 
The next step is to solve the zero-mode equation \eqref{zero_mode_eq} which,  for the field configuration \eqref{field_config_3d}, can be written as 
\begin{equation}
\partial^{2}\omega^{a}+A^{b}_{\mu}A^{b}_{\mu}\omega^{a}-A^{a}_{\mu}A^{b}_{\mu}\omega^{b}=0\,.
\end{equation}
It gives rise to the following system of equations:
\begin{eqnarray}
\partial^{2}\omega_{1}+r^{2}g^{2}(r)(\omega_{1}-\sin^{2}\theta\,\sin^{2}\varphi\,\omega_{1}
+\sin^{2}\theta\,\sin\varphi\,\cos\varphi\,\omega_{2})&=&0\,,\nonumber\\
\partial^{2}\omega_{2}+r^{2}g^{2}(r)(\omega_{2}-\sin^{2}\theta\,\cos^{2}\varphi\,\omega_{2}
+\sin^{2}\theta\,\sin\varphi\,\cos\varphi\,\omega_{1})&=&0\,,
\end{eqnarray}
where we have employed spherical coordinates 
\begin{eqnarray}
x&=&r\cos\varphi\,\sin\theta\,,\nonumber\\
y&=&r\sin\varphi\,\sin\theta\,,\nonumber\\
z&=&r\cos\theta\,. \label{3sc}
\end{eqnarray}
Parametrizing $(\omega_{1},\omega_{2})$ as 
\begin{eqnarray}
\omega_{1}&=&\sigma(r,\theta)\cos\varphi\,,\nonumber\\
\omega_{2}&=&\sigma(r,\theta)\sin\varphi\,,
\end{eqnarray}
we obtain
\begin{equation}
\frac{1}{r^{2}}\frac{\partial}{\partial r}\left(r^{2}\frac{\partial\sigma}{\partial r}\right)
+\frac{1}{r^{2}\sin\theta}\frac{\partial}{\partial\theta}\left(\sin\theta\frac{\partial\sigma}{\partial\theta}\right)
-\frac{\sigma}{r^{2}\sin^{2}\theta}+r^{2}g^{2}(r)\sigma=0\,.  \label{g3}
\end{equation}
Setting
\begin{equation}
\sigma(r,\theta)=r\psi(r)\xi(\theta)\,,
\end{equation}
and using the method of separation of variables, eq.\eqref{g3} splits into the two equations 
\begin{equation}
\frac{d}{d\theta}\left(\sin\theta\frac{d\xi(\theta)}{d\theta}\right)
-\frac{\xi(\theta)}{\sin\theta}
+k\sin\theta\,\xi(\theta)=0\,,  \label{k1}
\end{equation}
and 
\begin{equation}
r^{2}\psi''(r)+4r\psi'(r)+(2-k)\psi(r)+r^{2}g^{2}(r)\psi(r)=0\,, \label{psixieqs} 
\end{equation}
where $k$ is a constant parameter. \\\\Equation  \eqref{k1} can be written as
\begin{equation}
\frac{d^{2}\xi}{d\theta^{2}}+\cot\theta\,\frac{d\xi}{d\theta}
+\left(k-\frac{1}{\sin^{2}\theta}\right)\xi=0\,.
\end{equation}
Its solutions can be given in terms of associated Legendre polynomials $P^{m}_{\ell}(\cos\theta)$ with $k\equiv\ell(\ell+1)\geq2$ and
$m=1$, {\it i.e.}
\begin{equation}
\xi_{k=\ell(\ell+1)}(\theta)=P^{1}_{\ell}(\cos\theta)\,,
\end{equation}
with
\begin{equation}
\ell=1,2,3,4\dots\,,\qquad\mathrm{and}\qquad
k=\ell(\ell+1)=2,6,12,20,\dots\,.
\end{equation}
Turning now to eq.\eqref{psixieqs}, we follow Henyey's construction \cite{Henyey:1978qd} and proceed as in the previous section, {\it i.e.} eq.\eqref{psixieqs} is employed as an algebraic equation to express the function $g(r)$ in terms of $\psi(r)$, yielding 
\begin{equation}
r^{2}g^{2}(r)=-\frac{1}{r^{2}\psi(r)}\,\Bigl[r^{2}\psi''(r)
+4r\psi'(r)
+(2-k)\psi(r)\Bigr]\,.\label{r2g2}  
\end{equation}
As $g(r)$ is arbitrary, we need to propose a solution for $\psi(r)$ in such a way that the quantity $r^{2}g^{2}(r)$ is positive and, consequently, $A^{a}_{\mu}$ are real. A suitable choice for $\psi(r)$ is
\begin{equation}
\psi(r)=\frac{1}{r^{p}+\beta}\,,\label{psiguess}
\end{equation}
with $p>5/2$ in order to have a normalizable zero-mode. Indeed substituting \eqref{psiguess} in \eqref{r2g2} we get:
\begin{equation}
r^{2}g^{2}(r)=\frac{1}{r^{2}(r^{p}+\beta)^{2}}\,\Bigl[(3p-p^{2}+k-2)r^{2p}
+(3p+p^{2}+2k-4)\beta r^{p}
+(k-2)\beta^{2}\Bigr]\,, 
\end{equation}
which turns out to be positive for 
\begin{equation}
k\geq2\,,\qquad p>5/2\,,\qquad  {\rm and} \qquad (3-p)p+k-2\geq0\,.
\end{equation}
Let us check the boundary conditions, beginning with the evaluation of the Yang-Mills action. For the field configuration \eqref{field_config_3d},  the components of the field strength are given by
\begin{eqnarray}
F_{\mu\nu}&=&(\e_{1\mu\s}\,\e_{2\nu\rho}-\e_{2\mu\s}\,\e_{1\nu\rho})\,x_{\s}x_{\rho}\,g^{2}(r)\,,\nonumber\\
F^{a}_{\mu\nu}&=&-2\e_{a\mu\nu}\,g(r)+\frac{g'(r)}{r}\,\e_{a\rho\s}\,x_{\s}x_{\lambda}(\d_{\mu\lambda}\d_{\nu\rho}-
\d_{\mu\rho}\d_{\nu\lambda})\,.
\end{eqnarray}
Therefore 
\begin{eqnarray}
S_{\mathrm{YM}}&=&\frac{1}{4g^{2}}\int d^{3}x\,\left(F^{a}_{\mu\nu}F^{a}_{\mu\nu}+F_{\mu\nu}F_{\mu\nu}\right)\nonumber\\
&=&\frac{\pi}{g^{2}}\int_{0}^{\infty}dr\,\left[16\,r^{2}g^{2}(r)+\frac{8}{3}\,r^{3}g(r)g'(r)+r^{6}g^{4}(r)
+\frac{32}{15}\,r^{4}{g'}^{2}(r)\right] < \infty \;,
\end{eqnarray}
which is finite since $g(r)\sim1/r^{2}$ and $g'(r)\sim1/r^{3}$ as $r\to\infty$. \\\\Concerning now the Hilbert norm, one finds
\begin{eqnarray}
\|\mathcal{A}\|^{2}&=&\frac{16\pi}{3}\int_{0}^{\infty}dr\,r^{4}g^{2}(r)\nonumber\\
&=&\frac{16\pi}{3}\int_{0}^{\infty}\frac{dr}{(r^{p}+\beta)^{2}}\,\Bigl[(3p-p^{2}+k-2)r^{2p}
+(3p+p^{2}+2k-4)\beta r^{p}
+(k-2)\beta^{2}\Bigr]\,.
\end{eqnarray}
In general, since $g(r)\sim1/r^{2}$ as $r\to\infty$, the integral above is linearly divergent. However, we notice that there exists 
a particular value of $p$ and $k$ that makes this integral finite. Let  us consider in fact  the  case in which 
\begin{equation}
(3-p)p+k-2=0\,. 
\end{equation}
A solution of this equation is, for example,   $p=3$ and $k=2$, yielding a  Hilbert norm which is convergent, namely 
\begin{equation}
\|\mathcal{A}\|^{2}\Bigl|_{p=3,k=2}=96\pi\beta\int_{0}^{\infty}dr\,\frac{r^{3}}{(r^{3}+\beta)^{2}} <\infty\,.
\end{equation}

\section{The $d=4$ case}

Let us now face the case of $d=4$. Employing polar coordinates
\begin{eqnarray}
x&=&r\cos\varphi\,\sin\theta\,\sin\alpha\,,\nonumber\\
y&=&r\sin\varphi\,\sin\theta\,\sin\alpha\,,\nonumber\\
z&=&r\cos\theta\,\sin\alpha\,,\nonumber\\
t&=&r\cos\alpha\,,
\end{eqnarray}
where
\begin{equation}
r\geq0\,,\qquad0\leq\varphi\leq2\pi\,,\qquad0\leq\theta\leq\pi\,,\qquad0\leq\alpha\leq\pi\,, 
\end{equation}
for the Laplacian operator we have
\begin{equation}
\partial^{2}=\frac{1}{r^{3}}\frac{\partial}{\partial r}\biggl(r^{3}\frac{\partial }{\partial r}\biggr)
+\frac{1}{r^{2}\sin^{2}\alpha}\frac{\partial}{\partial\alpha}\biggl(\sin^{2}\alpha\,\frac{\partial }{\partial\alpha}\biggl)
+\frac{1}{r^{2}\sin^{2}\alpha}\left[\frac{1}{\sin\theta}\frac{\partial}{\partial\theta}\biggl(\sin\theta\frac{\partial }{\partial\theta}\biggr)+\frac{1}{\sin^{2}\theta}\frac{\partial^{2}}{\partial\varphi^{2}}\right]\,.
\end{equation}
In this case, as starting field configuration we  shall choose
\begin{equation}
A_{\mu}=0\,,\qquad A^{a}_{\mu}=\varepsilon_{a\mu\nu4}\,x_\nu\,\frac{h(r)}{\sin\alpha}\,,
\label{field_config_4d}
\end{equation}
where $h(r)$ is, as in the previous cases, a spherical symmetric function. Let us write the off-diagonal components in an explicit way, in order to show that they are not singular at $\alpha=0$, {\it i.e.}
\begin{eqnarray}
A^{a}_{4}&\!\!\!\!=\!\!\!\!&0\,,\nonumber\\
A^{1}_{1}&\!\!\!\!=\!\!\!\!&A^{2}_{2}=A^{3}_{3}=0\,,\nonumber\\
A^{1}_{2}&\!\!\!\!=\!\!\!\!&-A^{2}_{1}\,\,=\,\,z\,\frac{h(r)}{\sin\alpha}=rh(r)\,\cos\theta\,,\nonumber\\
A^{1}_{3}&\!\!\!\!=\!\!\!\!&-A^{3}_{1}=-y\frac{h(r)}{\sin\alpha}=-rh(r)\sin\varphi\sin\theta\,,\nonumber\\
A^{2}_{3}&\!\!\!\!=\!\!\!\!&-A^{3}_{2}=x\,\frac{h(r)}{\sin\alpha}=rh(r)\cos\varphi\sin\theta\,.
\end{eqnarray}
As $A_{\mu}=0$, we have to prove that $\partial_{\mu}A^{a}_{\mu}=0$ in order to check that the gauge conditions, eq.\eqref{gauge_fixing}, are fulfilled. First we notice that
\begin{equation}
x^{2}+y^{2}+z^{2}=r^{2}\sin^{2}\alpha\,,
\end{equation}
and
\begin{equation}
\sin\alpha=\frac{1}{r}\sqrt{x^{2}+y^{2}+z^{2}}=\frac{1}{r}\sqrt{r^{2}-t^{2}}\,. 
\end{equation}
Thus
\begin{equation}
A^{a}_{\mu}=\varepsilon_{a\mu\nu4}\,x_{\nu}\,\frac{rh(r)}{\sqrt{r^{2}-t^{2}}}\,.
\end{equation}
Now, acting with $\partial_{\mu}$ on $A^{a}_{\mu}$, we obtain
\begin{equation}
\partial_{\mu}A^{a}_{\mu}=\underbrace{\varepsilon_{a\mu\nu4}\,\delta_{\mu\nu}\,\frac{rh(r)}{\sqrt{r^{2}-t^{2}}}}_{=0}
+\varepsilon_{a\mu\nu4}\,x_{\nu}\,\partial_{\mu}\biggr(\frac{rh(r)}{\sqrt{r^{2}-t^{2}}}\biggl) \;.
\end{equation}
Since  $A^{a}_{4}=0$, we have to show that $\partial_{\mu}A^{a}_{\mu}=0$ for $\mu=1,2,3$. In fact, it is possible to prove that
\begin{equation}
\partial_{\mu}\biggr(\frac{rh(r)}{\sqrt{r^{2}-t^{2}}}\biggl)
=-x_{\mu}\left[\frac{rh(r)}{(r^{2}-t^{2})^{3/2}}
-\frac{1}{r\sqrt{r^{2}-t^{2}}}\frac{d}{dr}(rh(r))\right]\,,\qquad \mu=1,2,3\,,
\end{equation}
so that 
\begin{equation}
\partial_{\mu}A^{a}_{\mu}=-\varepsilon_{a\mu\nu4}\,x_{\mu}x_{\nu}\left[\frac{rh(r)}{(r^{2}-t^{2})^{3/2}}
-\frac{1}{r\sqrt{r^{2}-t^{2}}}\frac{d}{dr}(rh(r))\right]=0\,,\qquad \mu=1,2,3\,.
\end{equation}
Recalling now that $A_{\mu}=0$,  the zero-mode equation assumes the simplest form
\begin{equation}
\partial^{2}\omega^{a}+A^{b}_{\mu}A^{b}_{\mu}\omega^{a}-A^{a}_{\mu}A^{b}_{\mu}\omega^{b}=0\,.
\end{equation}
Let us see how the term $A^{b}_{\mu}A^{b}_{\mu}$ can be written by making use of  the particular configuration we have chosen, eq.\eqref{field_config_4d},  
\begin{eqnarray}
A^{a}_{\mu}A^{a}_{\mu}&=&A^{1}_{\mu}A^{1}_{\mu}+A^{2}_{\mu}A^{2}_{\mu}\nonumber\\
&=&A^{1}_{2}A^{1}_{2}+A^{1}_{3}A^{1}_{3}+A^{2}_{1}A^{2}_{1}+A^{2}_{3}A^{2}_{3}\nonumber\\
&=&(z^{2}+y^{2}+z^{2}+x^{2})\frac{h^{2}(r)}{\sin^{2}\alpha}\nonumber\\
&=&\frac{h^{2}(r)}{\sin^{2}\alpha}(r^{2}+z^{2}-t^{2})\nonumber\\
&=&\frac{r^{2}h^{2}(r)}{\sin^{2}\alpha}(1+\cos^{2}\theta\sin^{2}\alpha-\cos^{2}\alpha)\,.
\end{eqnarray}
Thus, we have the system of equations
\begin{eqnarray}
\partial^{2}\omega_{1}+\frac{h^{2}(r)}{\sin^{2}\alpha}(r^{2}+z^{2}-t^{2})\omega_{1}-A^{1}_{\mu}A^{1}_{\mu}\omega_{1}
-A^{1}_{\mu}A^{2}_{\mu}\omega_{2}&=&0\,,\nonumber\\
\partial^{2}\omega_{2}+\frac{h^{2}(r)}{\sin^{2}\alpha}(r^{2}+z^{2}-t^{2})\omega_{2}-A^{2}_{\mu}A^{1}_{\mu}\omega_{1}
-A^{2}_{\mu}A^{2}_{\mu}\omega_{2}&=&0\,.
\end{eqnarray}
Moreover, from 
\begin{eqnarray}
A^{1}_{\mu}A^{1}_{\mu}&=&A^{1}_{2}A^{1}_{2}+A^{1}_{3}A^{1}_{3}=\frac{h^{2}(r)}{\sin^{2}\alpha}(r^{2}-x^{2}-t^{2})\,,\nonumber\\
A^{2}_{\mu}A^{2}_{\mu}&=&A^{2}_{1}A^{2}_{1}+A^{2}_{3}A^{2}_{3}=\frac{h^{2}(r)}{\sin^{2}\alpha}(r^{2}-y^{2}-t^{2})\,,\nonumber\\
A^{1}_{\mu}A^{2}_{\mu}&=&A^{1}_{3}A^{2}_{3}=-\frac{h^{2}(r)}{\sin^{2}\alpha}\,xy\,,
\end{eqnarray}
it follows 
\begin{eqnarray}
\partial^{2}\omega_{1}+\frac{h^{2}(r)}{\sin^{2}\alpha}(r^{2}-y^{2}-t^{2})\omega_{1}+\frac{h^{2}(r)}{\sin^{2}\alpha}\,xy\,\omega_{2}&=&0\,,\nonumber\\
\partial^{2}\omega_{2}+\frac{h^{2}(r)}{\sin^{2}\alpha}(r^{2}-x^{2}-t^{2})\omega_{2}+\frac{h^{2}(r)}{\sin^{2}\alpha}\,xy\,\omega_{1}&=&0\,,
\end{eqnarray}
or, in spherical coordinates,
\begin{eqnarray}
\partial^{2}\omega_{1}+r^{2}h^{2}(r)\left(\omega_{1}-\sin^{2}\varphi\,\sin^{2}\theta\,\omega_{1}+
\cos\varphi\,\sin\varphi\,\sin^{2}\theta\,\omega_{2}\right)&=&0\,,\nonumber\\
\partial^{2}\omega_{2}+r^{2}h^{2}(r)\left(\omega_{2}-\cos^{2}\varphi\,\sin^{2}\theta\,\omega_{2}+
\cos\varphi\,\sin\varphi\,\sin^{2}\theta\,\omega_{1}\right)&=&0\,.
\end{eqnarray}
Setting
\begin{eqnarray}
\omega_{1}&=&\sigma(r,\theta,\alpha)\,\cos\varphi\,,\nonumber\\
\omega_{2}&=&\sigma(r,\theta,\alpha)\,\sin\varphi\,,
\end{eqnarray}
yields the following partial differential equation for  $\sigma(r,\theta,\alpha)$:
\begin{equation}
\left(\triangle-\frac{1}{r^{2}\sin^{2}\alpha\,\sin^{2}\theta}+r^{2}h^{2}(r)\right)\,\sigma(r,\theta,\alpha)=0\,, \label{dfeq4}
\end{equation}
where the differential operator $\triangle$ is given by
\begin{eqnarray}
\triangle&:=&\partial^{2}-\frac{1}{r^{2}\sin^{2}\alpha\,\sin^{2}\theta}\,\frac{\partial^{2}}{\partial\varphi^{2}}\nonumber\\
&=&\frac{1}{r^{3}}\frac{\partial}{\partial r}\biggl(r^{3}\frac{\partial }{\partial r}\biggr)
+\frac{1}{r^{2}\sin^{2}\alpha}\frac{\partial}{\partial\alpha}\biggl(\sin^{2}\alpha\,\frac{\partial }{\partial\alpha}\biggl)
+\frac{1}{r^{2}\sin^{2}\alpha\sin\theta}\frac{\partial}{\partial\theta}\biggl(\sin\theta\frac{\partial }{\partial\theta}\biggr)\,.
\end{eqnarray}
As done in the previous sections,  we try to solve the differential equation \eqref{dfeq4} by using the  method of the separation of variables and write $\sigma(r,\theta,\alpha)$ as
\begin{equation}
\sigma(r,\theta,\alpha)=R(r)\eta(\alpha)\xi(\theta)\,.
\end{equation}
In terms of these functions, the differential equation \eqref{dfeq4}  can be written as
\begin{eqnarray}
&&\frac{\eta\xi}{r^{3}}\frac{d}{dr}\biggl(r^{3}\frac{dR}{dr}\biggr)
+\frac{R\xi}{r^{2}\sin^{2}\alpha}\frac{d}{d\alpha}\biggl(\sin^{2}\alpha\frac{d\eta}{d\alpha}\biggr)
+\frac{R\eta}{r^{2}\sin^{2}\alpha\,\sin\theta}\frac{d}{d\theta}\biggl(\sin\theta\frac{d\xi}{d\theta}\biggr)\nonumber\\
&&-\frac{R\eta\xi}{r^{2}\sin^{2}\alpha\,\sin^{2}\theta}
+r^{2}h^{2}(r)\,R\eta\xi=0\,,
\end{eqnarray}
which, after multiplying by $r^{2}/R\eta\xi$, becomes
\begin{eqnarray}
&&\left[\frac{1}{rR}\frac{d}{dr}\biggr(r^{3}\frac{dR}{dr}\biggl)+r^{4}h^{2}(r)\right]\nonumber\\
&&+\left[\frac{1}{\eta\sin^{2}\alpha}\frac{d}{d\alpha}\biggl(\sin^{2}\alpha\frac{d\eta}{d\alpha}\biggr)
+\frac{1}{\xi\sin^{2}\alpha\,\sin\theta}\frac{d}{d\theta}\biggl(\sin\theta\frac{d\xi}{d\theta}\biggr)
-\frac{1}{\sin^{2}\alpha\sin^{2}\theta}\right]=0\,.
\end{eqnarray}
This equation holds true only if
\begin{equation}
\frac{1}{rR}\frac{d}{dr}\biggr(r^{3}\frac{dR}{dr}\biggl)+r^{4}h^{2}(r)=k\,,
\label{radeq}
\end{equation}
\begin{equation}
\frac{1}{\eta\sin^{2}\alpha}\frac{d}{d\alpha}\biggl(\sin^{2}\alpha\frac{d\eta}{d\alpha}\biggr)
+\frac{1}{\xi\sin^{2}\alpha\,\sin\theta}\frac{d}{d\theta}\biggl(\sin\theta\frac{d\xi}{d\theta}\biggr)
-\frac{1}{\sin^{2}\alpha\sin^{2}\theta}=-k\,,\label{ang_sec}
\end{equation}
where $k$ is a constant parameter.
\subsection{The angular sector}
Let us consider first equation \eqref{ang_sec}, which  can be rewritten as
\begin{equation}
\left[\frac{1}{\eta(\alpha)}\frac{d}{d\alpha}\biggr(\sin^{2}\alpha\,\frac{d\eta(\alpha)}{d\alpha}\biggl)+k\sin^{2}\alpha\right]
+\left[\frac{1}{\xi(\theta)\sin\theta}\frac{d}{d\theta}\biggr(\sin\theta\,\frac{d\xi(\theta)}{d\theta}\biggl)
-\frac{1}{\sin^{2}\theta}\right]=0\,,
\end{equation}
implying that
\begin{equation}
\frac{1}{\eta(\alpha)}\frac{d}{d\alpha}\biggr(\sin^{2}\alpha\,\frac{d\eta(\alpha)}{d\alpha}\biggl)+k\sin^{2}\alpha=\gamma\,,
\label{alphaeq}
\end{equation}
and
\begin{equation}
\frac{1}{\xi(\theta)\sin\theta}\frac{d}{d\theta}\biggr(\sin\theta\,\frac{d\xi(\theta)}{d\theta}\biggl)
-\frac{1}{\sin^{2}\theta}=-\gamma\,,
\label{thetaeq}
\end{equation}
where $\gamma$ is constant parameter. Moreover, eq.\eqref{thetaeq} can be cast in the form
\begin{equation}
\frac{d^{2}\xi(\theta)}{d\theta^{2}}+\cot\theta\,\frac{d\xi(\theta)}{d\theta}
+\left(\gamma-\frac{1}{\sin^{2}\theta}\right)\xi(\theta)=0\,,
\end{equation}
whose solution is given in terms of the associated Legendre polynomials $P^{m}_{\ell}(\cos\theta)$, with $\gamma\equiv\ell(\ell+1)\geq2$ and $m=1$, {\it i.e.}
\begin{equation}
\xi_{\gamma=\ell(\ell+1)}(\theta)=P^{1}_{\ell}(\cos\theta)\,,
\end{equation}
where
\begin{equation}
\ell=1,2,3,4\dots\,,\qquad\mathrm{and}\qquad\gamma=\ell(\ell+1)=2,6,12,20,\dots\,.
\label{gamma_values}
\end{equation}
We turn now to eq.\eqref{alphaeq} which takes the form
\begin{equation}
\sin^{2}\alpha\,\frac{d^{2}\eta(\alpha)}{d\alpha^{2}}+2\sin\alpha\,\cos\alpha\,\frac{d\eta(\alpha)}{d\alpha}
+(k\sin^{2}\alpha-\gamma)\eta(\alpha)=0\,.\label{mortaleq}
\end{equation}
In order to deal with this differential equation is convenient to perform the following change of variables
\begin{equation}
\sin\alpha=s\,,\qquad \cos\alpha=\pm\sqrt{1-s^{2}}\,,
\end{equation}
and define
\begin{equation}
\tau(s)=\eta(\alpha)\,.
\end{equation}
Then, we have
\begin{eqnarray}
\frac{d\eta(\alpha)}{d\alpha}&=&\pm\sqrt{1-s^{2}}\,\,\frac{d\tau(s)}{ds}\,,\nonumber\\
\frac{d^{2}\eta(\alpha)}{d\alpha^{2}}&=&(1-s^{2})\frac{d^{2}\tau(s)}{ds^{2}}
-s\,\frac{d\tau(s)}{ds}\,,
\end{eqnarray}
and  eq.\eqref{mortaleq} is replaced by 
\begin{equation}
s^{2}(1-s^{2})\frac{d^{2}\tau(s)}{ds^{2}}+s(2-3s^{2})\frac{d\tau(s)}{ds}+(ks^{2}-\gamma)\tau(s)=0\,.
\label{s_eq}
\end{equation}
We are interested in obtaining only a particular solution of the differential equation above, a task which is not too difficult. Indeed, a particular solution of \eqref{s_eq} is provided by 
\begin{equation}
\tau_{v}(s)=s^{v}\,,
\end{equation}
with $v$ being a real number, if, and only if,  the parameters $k$ and $\gamma$ are given by:
\begin{equation}
k=v(v+2)\,,\qquad \gamma=v(v+1)\,.
\end{equation}
As $\gamma$ is already fixed according to  eq.\eqref{gamma_values}, we conclude that $v$ has to be an integer number, just like the parameter $\ell$ of the Legendre polynomials. In this way,  the possible values of $\gamma$, fixed by eq.\eqref{gamma_values}, will be recovered. Some possible values of $v$ and  corresponding values of $\gamma$ and $k$ are displayed below:
\begin{equation}
\begin{tabular}{c|c|c}
$v$&$k$&$\gamma$\cr
\hline
$-3$&$3$&$6$\cr
$-2$&$0$&$2$\cr
$1$&$3$&$2$\cr
$2$&$8$&$6$\cr
$3$&$15$&$12$\cr
\end{tabular}
\end{equation}
Notice that  $v=-1$ is not possible because it would give $\gamma=0$, which is forbidden by eq.\eqref{gamma_values}. Also, the value $v=0$ does not give a solution of eq.\eqref{s_eq}. Notice that we have not fixed any particular value of $k$. In the next section, when we shall deal with the radial equation, eq.\eqref{radeq}, we shall impose some conditions for $k$ in order to obtain a real solution, as we have already done in the  $d=2$ and $d=3$ cases.  Another  point to be considered here is that we are looking for a normalizable zero-mode. In particular, this means that the angular integral
\begin{equation}
\int_{0}^{\pi}d\alpha\,\sin^{2}\alpha\, \tau_{v}^{2}(\sin\alpha)= \int_{0}^{\pi}d\alpha\,(\sin\alpha)^{2+2v}\,,  \label{ec}
\end{equation}
has to converge. For negative values of $v$ this integral may not converge, but for positive values of $v$, {\it e.g.} for $v=1$, the integral does converge. In fact, taking $v=1$ we obtain,
\begin{equation}
\int_{0}^{\pi}d\alpha\,(\sin\alpha)^{4}=\frac{3\pi}{8}\,.
\end{equation}
From now on we shall resrict to positive values of $v$, for which expression \eqref{ec} is convergent. 

\subsection{The radial sector}
Let us look now at the radial equation \eqref{radeq}, which can be written as
\begin{equation}
\frac{1}{r}\frac{d}{dr}\biggr(r^{3}\frac{dR(r)}{dr}\biggl)
+r^{4}h^{2}(r)R(r)-kR(r)=0\,.
\end{equation}
Setting
\begin{equation}
R = r\psi(r)\,,
\end{equation}
we get
\begin{equation}
r^{2}h^{2}(r)=-\frac{1}{r^{2}\psi(r)}\Bigl[r^{2}\psi''(r)+5r\psi'(r)+(3-k)\psi(r)\Bigr]\,.\label{real}
\end{equation}
As before, our first task is to show that the quantity above is positive for a given $\psi(r)$. This will ensure that the gauge field components $A^{a}_{\mu}$ are real. As usual, we shall  look at  $\psi(r)$ in the form
\begin{equation}
\psi(r)=\frac{1}{r^{p}+\beta}\,,\label{psi}
\end{equation}
where $\beta$ is a positive constant parameter. In this case, we have to require that $p>3$ in order to have a normalizable zero-mode, in fact 
\begin{equation}
\|\omega\|^{2}\propto \int_{0}^{\infty}r^{3}dr\frac{r^{2}}{(r^{p}+\beta)^{2}}\;
<\infty\,,  \qquad {\rm for } \;\;\;\;p>3 \;. 
\end{equation}
Plugging expression \eqref{psi} into eq.\eqref{real}, we obtain
\begin{equation}
r^{2}h^{2}(r)=\frac{1}{r^{2}(r^{p}+\beta)^{2}}\Bigl[(4p-p^{2}+k-3)r^{2p}
+(p^{2}+4p+2k-6)\beta r^{p}-(3-k)\beta^{2}\Bigr]\,,  \label{hr}
\end{equation}
which is always positive if
\begin{equation}
4p-p^{2}+k-3\geq0\,,\qquad\mathrm{and}\qquad k\geq3\,.
\end{equation}
Collecting all expressions above, we can describe the  zero-mode solution with only  two parameters $(p,v)$, which are constrained by the following conditions:
\begin{equation}
p>3\,,\qquad v=1,2,3,4,\dots\,,\qquad\mathrm{and}\qquad
4p-p^{2}+v(v+2)-3\geq0\,,
\end{equation}
where use has been made of the relation $k=v(v+2)$. Thus, the components of the zero-mode read
\begin{eqnarray}
\omega_{1}(r,\alpha,\theta,\varphi)&=&\frac{r}{(r+\beta)^{p}}(\sin\alpha)^{v}P^{1}_{v}(\cos\theta)\cos\varphi\,,\nonumber\\
\omega_{2}(r,\alpha,\theta,\varphi)&=&\frac{r}{(r+\beta)^{p}}(\sin\alpha)^{v}P^{1}_{v}(\cos\theta)\sin\varphi\,.
\end{eqnarray}
As done in the previous sections, we check now the boundary conditions. For the field configuration \eqref{field_config_4d}  the components of the field strength are given by:
\begin{equation}
F_{\mu\nu}=\e^{ab}\,A^{a}_{\mu}A^{b}_{\nu}\,,\qquad
F^{a}_{\mu\nu}=\partial_{\mu}A^{a}_{\nu}-\partial_{\nu}A^{a}_{\mu}\,,
\end{equation}
as it follows by noticing that the diagonal configuration vanishes, $A_\mu=0$.  The Yang-Mills action is then given by
\begin{eqnarray}
S_{\mathrm{YM}}&=&\frac{1}{4g^{2}}\int d^{4}x\,\left(F^{a}_{\mu\nu}F^{a}_{\mu\nu}+F_{\mu\nu}F_{\mu\nu}\right)\nonumber\\
&=&\frac{\pi^{2}}{g^{2}}\int_{0}^{\infty}dr\,r^{3}\left(
\frac{1}{3}\,r^{4}h^{4}(r)+\frac{49}{30} h^{2}(r)-\frac{1}{5}\,rh(r)h'(r)
+\frac{49}{30}\,r^{2}{h'}^{2}(r)\right)\,.   \label{int4}
\end{eqnarray}
From expression \eqref{hr} it is easily seen that, for generic values of the parameters $(p,v)$, the function $h(r)$ behaves as $h(r)\sim1/r^{2}$  as $r\to\infty$. As such, the integral \eqref{int4} is  logarithmic  divergent. Nevertheless,  we can choose the parameters $p$ and $v$ in such a way that 
$h(r)$ behaves like $1/r^{n}$, with $n>2$, as $r\to \infty$. In fact, for $p=4$ and $v=1$ we obtain that $h(r)\sim1/r^{4}$ as $r\to \infty$, and expression  \eqref{int4} is finite. \\\\Concerning now the Hilbert norm   $\|\mathcal{A}\|^2$, we get 
\begin{eqnarray}
\|\mathcal{A}\|^{2}&=&\frac{8\pi^{2}}{3}\int_{0}^{\infty} dr\,r^{5}h^{2}(r)\,,
\end{eqnarray}
which is also convergent in the case: $p=4$, $v=1$.

\section{Conclusions}

In this work we have constructed classes of normalizable zero-modes for the Faddeev-Popov operator in  the maximal Abelian gauge  for  $SU(2)$ Yang-Mills gauge theory in $d=2,3,4$ Euclidean dimensions, generalizing previous works done mainly in the Landau gauge \cite{Guimaraes:2011sf, Capri:2012ev}. The  construction presented here has been  achieved by employing Henyey's method \cite{Henyey:1978qd}, which allows for a self-consistent evaluation of the corresponding gauge field configurations. \\\\Our results can be summarized as follows: 
\begin{itemize} 
\item for $d=2$ Euclidean dimensions we were able to obtain normalizable zero modes with corresponding gauge fields configurations displaying finite action but not  finite Hilbert  norm, 
\item  for $d=3$ and $d=4$ Euclidean dimensions we have obtained normalizable zero modes solutions with corresponding gauge fields configurations displaying finite action as well as finite norm $\|\mathcal{A}\|^2$. 
\end{itemize}

\section*{Acknowledgments}
The Conselho Nacional de Desesnvolvimento Cinet\'{\i}fico e Tecnol\'{o}gico (CNPq-Brazil), the FAPERJ, Funda\c{c}\~{a}o de Amparo \`{a} Pesquisa do Estado do Rio de Janeiro, the Coordena\c{c}\~{a}o de Aperfei\c{c}oamento Pessoal de N\'{\i}vel Superior (CAPES) are gratefully acknowledged.

\end{document}